\documentclass[twocolumn,pra,showpacks,preprintnumber]{revtex4} 
\usepackage[dvips]{graphicx}
\usepackage{dcolumn}
\usepackage{amsmath,amssymb}
\usepackage{txfonts}
\usepackage{bm}

\begin{document}
\draft
\title{Coherent Spin Dynamics into Space Quantization}
\author{Masato Morifuji}
\affiliation{Department of Electronic Engineering, Osaka University, 
Suita, Osaka 565-0871, Japan}
\date{\today}

\begin{abstract}
We propose a mechanism to describe how a physical quantity,
which initially can take continuous values, is restricted within 
some discrete values after a measurement.
As an example of the present theory, in which interplay between coherence 
of motion and fluctuation from disturbance plays an important role, 
we investigate motion of a spin state in a magnetic field.
First, we point out that discrete eigenstates are formed from 
continuous states as a result of coherence of precession motion of a spin.
Next, by assuming disturbance from environmental electromagnetic fields, 
we investigate temporal change of direction of a spin state 
by applying the first order perturbation theory and the Monte Carlo technique.
Results of simulations show that the spins, whose directions 
are randomly distributed at the initial time,  
are reoriented toward only two directions due to fluctuation 
guided by coherence.
\end{abstract}

\maketitle


\section{Introduction}

In spite of the great success in a large number of applications, 
the quantum mechanics still contains a puzzling aspect in its foundations.
One of the crucial questions is why and how an observed physical quantity is 
restricted within only some particular discrete values.
A typical example is seen in the historical experiment achieved 
by O. Stern and W. Gerlach in 1922.\cite{SG1,SG2,SG3}
As is well known, they investigated trajectories of Ag atoms 
injected into an inhomogeneous magnetic field from a hot oven 
for the purpose of measuring momentum of Ag atoms. 
Contrary to naive expectation they observed split atom beams.
This result indicated that magnetic momenta of the Ag atoms 
take only two values, 
even though the Ag atoms should have been randomly oriented 
when they escaped out of the hot oven.
This effect, called space quantization, clearly exhibited 
that physical phenomena of atomic scale are essentially different 
from the concept of classical mechanics.

To date, the quantization effect is the foundation of 
the quantum mechanics which is the only reliable theory to describe 
physics of atomic scale.
The quantization effect is often attributed to  
reduction of wavefunction which cannot be deduced from other principles.
It is sometimes said that a measurement always causes the system jumps into 
an eigenstate of the measured variable.
In spite of a number of theoretical trials to describe the reduction of a 
wavefunction in terms of a density matrix, the problem is 
still controversial.\cite{Namiki1,Namiki2,Joos,Zurek}

It has been long thought that such a problem is 
too academic and useless even if it is important as the foundation of 
the quantum mechanics.
However, recent development of technologies has enabled us to observe 
phenomena which could be carried out only in a conceptual experiment.
For example, some experiments of quantum information  have revealed essential 
aspects of the quantum mechanics as a real substance.\cite{QuantumInfo}
Behavior of electrons in a mesoscopic system also needs essential  
quantum mechanical viewpoints.\cite{Mesoscopic}
In addition, there are proposals of quantum computer utilizing nuclear 
spins\cite{Kane} or electron spins in a quantum dot.\cite{takeuchi}
Therefore, deeper insight for the quantum mechanics
is required, 
and it is of great significance to investigate motion of an spin 
for development of novel devices as well as for basic physics.

In this letter, we propose a mechanism to explain how a physical quantity 
is quantized into some discrete levels when an observation is carried out.
In order to describe the quantization process, 
we study the effect of quantum coherence on motion of a physical quantity
with an example of a spin in a magnetic field.
In the precedent studies, we have shown that coherence of time-evolving 
electron wave gives rise to formation of quantized eigenstates 
with discrete eigenenergies.\cite{Morifuji1,Morifuji2}
We have calculated time-dependent density of states which exhibits transformation 
from a continuous spectrum into discrete levels, however, we have not taken 
the effect of scattering into account.
In this study we investigate the effect of collaboration of coherence 
and fluctuation due to inelastic scattering, and show how 
a value of physical quantity approaches one of discrete eigenvalues. 

There are theories that dephasing due to interaction between a quantum and 
environment plays an important role in a measurement process.\cite{Joos,Zurek,Wezel}
The present study is based on the similar standpoint. However, as far as we know, 
there have been no studies that pointed out the importance of 
interplay of coherence and scattering.

\section{Theory}
\subsection{Formation of eigenstates due to coherence}

We investigate motion of a spin in a magnetic field, 
and show how it is quantized owing to interplay between coherence 
of motion and fluctuation due to environmental disturbance.

Let us consider a spin in a uniform magnetic field applied along the $z$-axis.
This is a similar situation to the Stern-Gerlach experiment (SGE). 
However, we note that the situation of this study is 
different from that of the SGE in some points.
First, we treat electron spins instead of momenta of Ag atoms.
Next, we consider a uniform magnetic field, instead of 
inhomogeneous field used in the SGE.
Inhomogeneity is only necessary to change electron's position 
in accordance with a direction of a spin. 
It is thus sufficient to consider motion in a uniform field when we are 
interested in directions of the spin.
It is not difficult to include spatial motion of an electron 
by using wavepacket wavefuctions.

We consider a spin which is oriented a $(\theta,\varphi)$ direction 
in the polar coordinate at the initial time $t =0$.
This spin state is expressed by a wavefunction 
\begin{align}
|\theta,\varphi\rangle = \cos(\theta/2) \,e^{-i\varphi/2}\, |+\rangle
                         + \sin(\theta/2) \,e^{i\varphi/2}\,  |-\rangle, 
\label{eq:eq1}
\end{align}
where $|+\rangle$ and $|-\rangle$ are the eigenstates of the spin operator $s_z$ as 
\begin{align}
s_{z}|\pm\rangle  = \pm \frac{\hbar}{2}\, |\pm\rangle.
\label{eq:eq2}
\end{align}
Since the Hamiltonian of the interaction between a spin and 
a magnetic field, ${\cal H} = (eB/m)s_{z}$, 
is a generator of a time-evolution operator, 
the spin state in a magnetic field shows precession around the $z$-axis 
with a frequency $\omega = eB/m$ as 
\begin{align}
|\theta,\varphi,t\rangle &= |\theta,\varphi+\omega t\rangle \notag\\
&= \cos(\theta/2) \,e^{-i(\varphi+\omega t)/2}\, |+\rangle
                    + \sin(\theta/2) \,e^{i(\varphi+\omega t)/2}\,  |-\rangle.
\label{eq:eq3}
\end{align}

We introduce here a superposition of the spin wavefunctions 
in the precessing motion as
\begin{align}
|\chi(\theta,\varphi,t)\rangle
 = \frac{1}{\sqrt{\,t\,}}\!
    \int_{0}^{\,t\,}\!\! dt' e^{iE_{\theta}t'/\hbar}\,|\theta,\varphi,t'\rangle.
\label{eq:eq4}
\end{align}
In this equation $E_{\theta}$ is energy of the spin state in a magnetic field 
given by 		
$E_{\theta}=\langle\theta,\varphi|{\cal H}|\theta,\varphi\rangle
=eB\hbar\cos\theta/2m$.
$|\chi(\theta,\varphi,t)\rangle$ is a coherent superposition of 
wavefunctions over the history during the time $0 \sim t$.
In other words, we may interpret the function $|\chi(\theta,\varphi,t)\rangle$ 
as afterimage of precessing spin. 
The function $|\chi(\theta,\varphi,t)\rangle$ has a physical meaning as 
a time-integrated probability amplitude, i.e., the probability of finding a
state at the direction $\theta$ during the period $0\sim t$.

\begin{figure}[h]
\includegraphics[width=5.0cm]{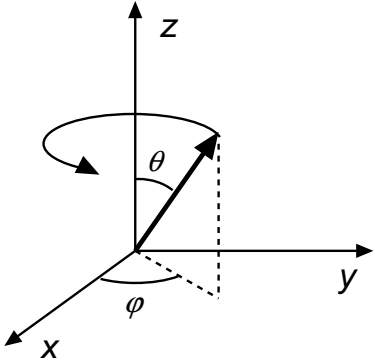}
\caption{Precession of a spin in a magnetic field applied along the $z$-axis.}
\label{fig:1}
\end{figure}

We note that the exponential factor in eq. (\ref{eq:eq4}) has a role to 
cancel the time-dependent phase factor of $|\theta,\varphi,t'\rangle$, i.e., 
$e^{-i E_{\theta}t'/\hbar}$.
This cancellation is important to obtain a rational result.
Since we are considering a superposition of states belonging to different times, 
relative phase difference between the states changes 
if additional constant potential is applied.
Thus the additional potential changes the superposed wavefunctions completely.
Since such a situation is nonsense, 
the time-dependent phase factor should be excluded from the time-integration 
 from the viewpoint of the gauge invariance of the superposed wavefunctions 
as shown in eq. (\ref{eq:eq4}).

By inserting eq. (\ref{eq:eq3}) into eq. (\ref{eq:eq4}), 
we can show that this function approaches the spin eigenstates with time as
\begin{align}
\lim_{t \to \infty} |\chi(\theta,\varphi,t)\rangle 
%
\propto
\begin{cases}
\ \  |+\rangle  \hspace{10mm} (\theta = 0 )      \\
\ \  |-\rangle  \hspace{10mm}(\theta = \pi )     \\
\ \  0          \hspace{12mm}({\rm others}). 
\end{cases}
\label{eq:eq5}
\end{align}
Eq. (5) means that we can interpret that eigenstates are formed as a result of coherence 
of motion of a precessing spin.
When energy of the precessing spin coincides with one of eigenenergies 
$E_{\pm}={\pm}eB\hbar/2m$, 
the coherent superposition approaches corresponding eigenstate.
On the other hand, if energy is apart from any of eigenenergies, 
superposition of moving states decays due to destructive self-interference.

\begin{figure}[h]
\includegraphics[width=6.5cm]{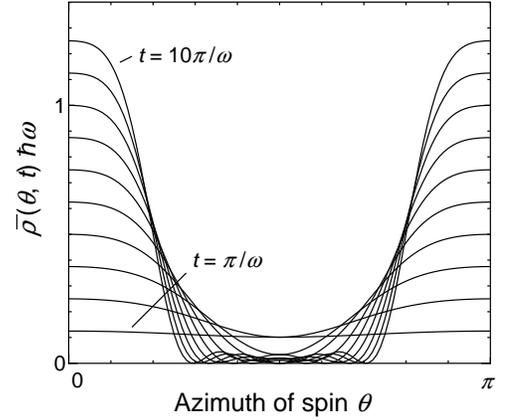}
\caption{The time-dependent density of states $\bar{\rho}(\theta,t)$ 
plotted as a function $\theta$. Eash curve shows $\bar{\rho}(\theta,t)$
for the time $t = \pi/\omega \sim 10\pi/\omega$.
Note that $\bar{\rho}(\theta,t)$ is normalized by a characteristic 
energy $\hbar\omega$.}
\label{fig:2}
\end{figure}

In order to show how the continuous spectrum changes into discrete levels, 
we introduce a function $\bar{\rho}(\theta,t)$ with a norm of 
the function $\chi(\theta,\varphi,t)$ given by 
\begin{align}
\bar{\rho}(\theta,t) 
&= \frac{1}{2\pi\hbar}\,
              \langle\chi(\theta,\varphi,t)\,|\,\chi(\theta,\varphi,t)\rangle \notag\\ 
&= \sum_{i=\pm} |A_{i}|^{2}\,\delta_{t}(E_{\theta} \mp \hbar\omega/2).
\label{eq:eq6}
\end{align}
In this equation, $A_{\pm} \equiv \langle\theta,\varphi,0|\pm\rangle$
 is an overlap integral between the initial state and the eigenstate.
$\delta_{t}(E)= \sin^{2}(Et/2\hbar)/(\pi E^{2}t/2\hbar)$ is the $\delta$-function 
broadened due to finite lifetime. 
Since $\mp \hbar\omega/2$ are eigenenergies in a magnetic field
we may interpret this function as density of states 
with broadened spectra due to finite lifetime 
formed as a result of coherent motion of a spin state.

Figure 2 shows $\bar{\rho}(\theta,t)$ normalized by a characteristic energy 
$\hbar\omega$ plotted as a function of $\theta$.
The curves show $\bar{\rho}(\theta,t)$ calculated for $t = \pi/\omega \sim 10\pi/\omega$.
When $t$ is small $\bar{\rho}(\theta,t)$ is an almost uniform function of $\theta$.
On the other hands, when $t$ is large, $\bar{\rho}(\theta,t)$ has peaks around 
$\theta = 0$ and $\pi$.
This means that probability to find the spin at the direction $\theta$ is 
restricted within the two values because of formation of eigenstates due to coherence.
Considering that $|\chi(\theta,\varphi,t)\rangle$ is relevant to observation
for finite duration of time, 
the behavior of $\bar{\rho}(\theta,t)$ means we find a spin at any directions 
when $t$ is small, whereas
 we find a spin only in the direction $\theta= 0$ or $\pi$ in the long time.

\subsection{Simulation of dynamics of spin direction}

Let us presume that the direction of a spin fluctuates due to 
disturbance from environmental electromagnetic fields.
Unfortunately, we lack detailed knowledge of the interaction 
between a spin and electromagnetic fields.
We thus treat strength of the interaction as a parameter, 
and apply the first order time-dependent perturbation theory 
to investigate fluctuation of spin directions 
with a conventional picture of sudden change of spin directions.

We have to note that the set of the states ${|\theta,\varphi\rangle}$'s is overcomplete.
In order to derive an expression of scattering probability 
with overcomplete basis including the effect of the coherence,
 we start from the Lippmann-Schwinger equation with the Born approximation
\begin{align}
|\psi^+\rangle = |\theta,\varphi\rangle -\frac{i}{\hbar}\!\int_{0}^{\,t}\!d\tau\, 
 e^{i(E_{\theta}-{\cal H})\tau/\hbar} V(\tau)\, |\theta,\varphi\rangle. 
\label{eq:eq7}
\end{align}
In this equation, $|\psi^+\rangle$ is a state 
generated from the initial state $|\theta,\varphi\rangle$  
due to interaction between a spin and electromagnetic fields. 
We assume that the interaction is given by the operator 
$V(\tau) = V\,e^{\pm i\Omega \tau/\hbar}$
with $\Omega$ the frequency of the electromagnetic field.
By inserting the completeness relation of the spin states
\begin{align}
\frac{1}{2\pi}\!\!\int_{0}^{\pi}\!\!d\theta'\!\!\int_{0}^{2\pi}\!\!\!d\varphi'\sin\theta'\,
|\theta',\varphi'\rangle\langle \theta',\varphi'| ={\bf 1}
\label{eq:eq8}
\end{align}
into eq. (\ref{eq:eq7}), we have
\begin{align}
|\psi^{+}\rangle &\simeq |\theta,\varphi\rangle 
    -\int_{0}^{\pi}\!\!d\theta'\!\!\int_{0}^{2\pi}\!\!\!d\varphi'\sin\theta' \notag \\
&\times\left[\frac{iV}{2\pi\hbar}    
\int_{0}^{\,t}\!d\tau  
\langle \theta',\varphi'|e^{i(E_{\theta}- {\cal H} \pm\hbar\Omega)\tau/\hbar} 
|\theta,\varphi\rangle \right]\, |\theta',\varphi'\rangle.
\label{eq:eq9}
\end{align}
This equation indicates that the scattered state $|\psi^{+}\rangle$ is written 
as a linear combination of spin states $|\theta',\varphi'\rangle$'s.
Therefore, by following the ordinary argument of the time-dependent perturbation theory, 
we have the expression of scattering probability as 
\begin{align}
W(\theta',\varphi'&;\theta,\varphi,t)
   \equiv  \frac{|a_{\theta',\varphi'}(t)|^{2}}{t}
  =  \frac{|V|^2 }{2\pi\hbar}\,\rho(\theta',t), 
\label{eq:eq10}
\end{align}
where $a_{\theta',\varphi'}(t)$ is the quantity in the square bracket 
in eq. ({\ref{eq:eq9}).
$\rho(\theta',t)$ in this equation is given by an expression 
similar to eq. (\ref{eq:eq6}) 
but with different factors
$B_{+} = \cos(\theta/2)\cos(\theta'/2)$ and $B_{-}=\sin(\theta/2)\sin(\theta'/2)$
instead of $A_{\pm}$. 
We may regard $\rho(\theta',t)$ as time-dependent density of states for transition.
Except the factor $2\pi$ arising from the prefactor of eq. (\ref{eq:eq8}), 
this expression is the same as the Fermi's golden rule.
See ref. \cite{Morifuji2} for the detail of scattering theory with non-orthogonal 
overcomplete basis.

We investigated temporal change of spin directions by using the Monte Carlo 
simulation, taking stochastic nature of behavior of electrons 
into account.\cite{Jacoboni}
The procedure of the simulation for a spin is 
\begin{description}
\item{(i)} We set an initial direction of a spin with a random number.
The spin begins precession.
\item{(ii)} By considering that precession motion is interrupted 
due to inelastic scattering after time $t_f$, 
we evaluate $t_f$ with an equation 
\begin{align}
-\log r_i = \int^{\,t_f}\!\!dt\, W(\theta,\varphi,t) 
\label{eq:eq11}
\end{align}
where $r_i$ is another random number and  
\begin{align}
W(\theta,\varphi,t) = 
\int_{0}^{\pi}\!\!d\theta'\!\!\int_{0}^{2\pi}\!\!\!d\varphi'\sin\theta'
W(\theta',\varphi';\theta,\varphi,t)
\label{eq:eq12}
\end{align}
is a total transition rate of the spin.
\item{(iii)} We determine a new direction of the spin using a random number 
with weight given by $W(\theta',\varphi';\theta,\varphi)$ which is evaluated 
from eq. (\ref{eq:eq10}). 
\item{(iv)} We repeat the steps (i), (ii), and (iii) 
considering that coherency is lost when inelastic scattering occurred, 
and accumulate relevant quantities.
\end{description}

\begin{figure}[h]
\includegraphics[width=7.0cm]{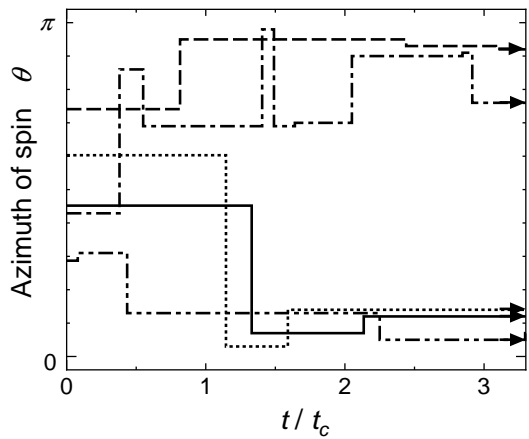}
\caption{Examples of spin motion calculated from the present theory.
Sample paths calculated by the Monte Carlo simulations are plotted as functions time.}
\label{fig:3}
\end{figure}
\begin{figure}[h]
\includegraphics[width=7.0cm]{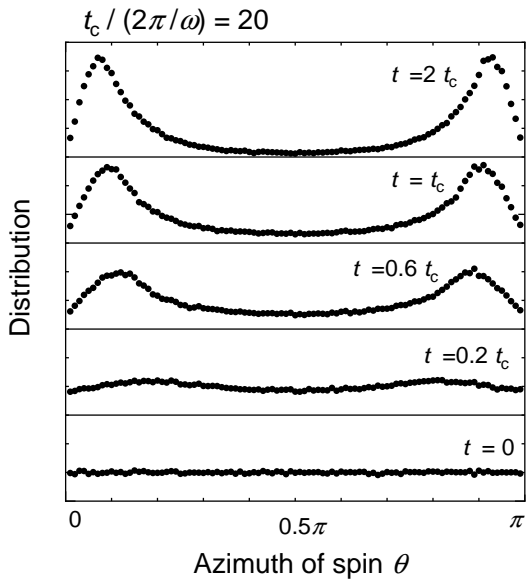}
\caption{Distribition of spin directions at the time $t= 0, 0.2t_c, 0.6t_c, t_c$ 
and $2t_c$ obtained from 100,000 paths of the Monte Carlo simulation 
with the parameter $t_c/(2\pi/\omega)=20$.}
\label{fig:4}
\end{figure}

\section{Results and Discussion}

Results of the Monte Carlo simulations are depicted in Figs. 3, 4 and 5.
In Fig. 3, some of the sample paths (temporal change of azimuth of spins) 
calculated for the parameter $t_c/(2\pi/\omega)=20$ 
are plotted as a  function of time.
The time is normalized by the characteristic time of scattering 
defined by $t_{c} \equiv 2\pi\hbar^{2}\omega/|V|^{2}$.
An average interval of scattering events may be represented by $t_c$.
As clearly shown in Fig. 3, the azimuth of spin fluctuates due to 
inelastic scattering, and 
approaches $\theta \simeq 0$ or $\pi$ after receiving several scattering 
indifferent to the initial direction.

Fig. 4 shows distribution of $\theta$ at the time $t= 0, 0.2t_c, 0.6t_c, t_c$ and $2t_c$
accumulated from 100,000 paths of the simulations.
At the initial time $t=0$, the azimuth distributes uniformly between $0$ and $\pi$.
With increasing time, the distribution curve shows two broad peaks
 corresponding to the most probable direction to be observed.
This is because, as also shown in Fig.3, a spin is reoriented toward 
the two directions $0$ or $\pi$ while receiving scattering several times.

\begin{figure}[h]
\includegraphics[width=6.5cm]{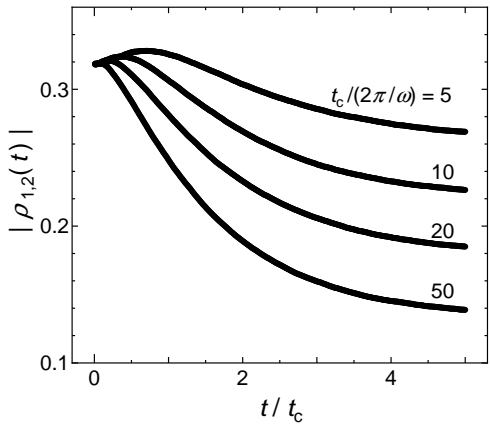}
\caption{The off-diagonal element $|\rho_{12}(t)|$ of the density matrix
 are plotted as a function of time.}
\label{fig:5}
\end{figure}

This quantization effect arises 
from interplay between coherence of precession motion and fluctuation 
due to inelastic scattering.
The transition probability is proportional to density of final states 
which is formed as a result of coherent superposition of moving spin states.
Since the density of states has peaks around $\theta = 0$ and $\pi$, 
the final states tend to have these values of azimuth.
This also means that there must be enough long time between scattering events 
for the coherence to be formed.
When an interval of successive scattering events is very long, possible final states 
of scattering will be restricted to $\theta = 0$ or $\pi$.
On the other hand, if scattering occurs very frequently, final states can have 
any directions.
In other words, it is necessary that the ratio between scattering interval and 
precession period $t_c/(2\pi/\omega)$ must be large for this mechanism of 
quantization to occur.

We can interpret the quantization process in terms of phase coherency 
of an ensemble of spins.
Properties of an ensemble is well expressed by a density matrix given 
by an average over $N$ spins as 
\begin{align}
\rho_{m,n}(t) = \frac{1}{N}\sum_{j=1}^{N}|\theta_j,\varphi_j,t\rangle
\langle\theta_j,\varphi_j,t|,
\end{align}
where $j$ specifies each spin.
Figure 5 shows the off-diagonal element of the density matrix evaluated 
from the simulations.
With increasing time, the off-diagonal element decreases,
indicating that the ensemble of spins approaches a mixed ensemble
as $\rho_{m,n} \simeq (1/2)\,\delta_{m,n}$ 
 due to reduction of phase coherency.

Finally we note that this theory can be verified in experiments.
Even though the strength of inelastic scattering is unknown,
we can vary the parameter $t/t_c$ by changing length of the region 
where a magnetic field exists.
It is also possible to change the parameter $t_c/(2\pi/\omega)$ 
by changing magnitude of the applied magnetic field. 
We expect that the distribution curves of spin direction as shown in Fig. 4 will be 
obtained from measurements with various conditions.

\section{Conclusion}

In conclusion, we have proposed a mechanism to describe 
how a physical quantity is quantized due to an observation.
We have shown that the interplay between coherence and fluctuation leads a spin 
toward one of eigenstates.
Base on this theory, 
we carried out simulations for motion of spin under both a magnetic field 
and disturbance from environment. 
The results of simulations clearly showed 
that the direction of the spin tends to approach only two values
$\theta=0$ or $\pi$ while receiving several scattering. 
The results elucidated that space quantization of a spin can be explained 
by the interplay between coherent motion and decoherence due to scattering.



\begin{references} 
\bibitem{SG1} W.~Gerlach and O.~Stern, Z. Phys. {\bf 9}, 349 (1922).
\bibitem{SG2} W.~Gerlach and O.~Stern, Z. Phys. {\bf 9}, 353 (1922).
\bibitem{SG3} W.~Gerlach and O.~Stern, Ann. d. Phys. {\bf 74}, 673 (1924).
\bibitem{Namiki1} M.~Namiki, S.~Pascazio and H.~Nakazato, {\it Decoherence and Quantum 
                  Measurements} (World Scientific, Singapore, 1997).
\bibitem{Namiki2} S.~Machida and M.~Namiki, Prog. Theor. Phys., 
                  {\bf 63}, 1457 (1980).
\bibitem{Joos} E.~Joos et al.,  {\it Decoherence and the Appearance of a 
               Classical World in Quantum Theory} 
               (Springer-Verlag, New York, 2003). 
\bibitem{Zurek} W.~H.~Zurek, Rev. Mod. Phys. {\bf 75} 715 (2003).
\bibitem{Wezel} J. van Wezel, J. van den Brink, and J. Zaanen, 
                Phys. Rev. Lett. {\bf 94} 230401 (2005).
\bibitem{QuantumInfo} N.~Takei, H.~Yonezawa, T.~Aoki, and A.~Furusawa, 
                      Phys. Rev. Lett. {\bf 94}, 220502 (2005).
\bibitem{Mesoscopic} Q.~Q.~Wang,A.~Muller, P.~Bianucci, E.~Rossi, Q.~K.~Xue, T.~Takagahara, 
                     C.~Piermarocchi, A,~H.~MacDonald, and C.~K.~Shih, 
                     Phys. Rev. B{\bf 72}, 035306 (2005).
\bibitem{Kane} B.~E.~Kane, Nature {\bf 393}, 133 (1998).
\bibitem{Takeuchi} A.~Takeuchi, T.~Kuroda, Y.~Nakata, M.~Murayama, T.~Kitamura, 
                   N.~Yokoyama, Jpn. J. Appl. Phys. {\bf 42}, 4278 (2003).
\bibitem{Morifuji1} M.~Morifuji and K.~Kato, Phys. Rev. B {\bf 68}, 035108 (2003).        
\bibitem{Morifuji2} M.~Morifuji, J. Phys. Soc. Jpn. {\bf 73}, 2174 (2004).
\bibitem{Jacoboni} C.~Jacoboni and L.~Reggiani, Rev. Mod. Phys. {\bf 55}, 645 (1983).
\end{references}
\end{document}